\documentstyle[11pt,aaspp4]{article}

\def\be{\begin{equation}}
\def\ee{\end{equation}}
\def\go{\mathrel{\raise.3ex\hbox{$>$}\mkern-14mu
             \lower0.6ex\hbox{$\sim$}}}
\def\lo{\mathrel{\raise.3ex\hbox{$<$}\mkern-14mu
             \lower0.6ex\hbox{$\sim$}}}

\begin{document}

\title{Transonic Magnetic Slim Accretion Disks and kilo-Hertz
Quasi-Periodic Oscillations in Low-Mass X-Ray Binaries}

\author{Dong Lai}
\affil{Theoretical Astrophysics, California Institute of
Technology, Pasadena, CA 91125;\\
Department of Astronomy, Space Sciences Building, Cornell University,
Ithaca, NY 14853;\\
E-mail: dong@spacenet.tn.cornell.edu}

\begin{abstract}
The inner regions of accretion disks of weakly magnetized 
neutron stars are affected by general relativistic gravity and
stellar magnetic fields. Even for field strengths sufficiently small
so that there is no well-defined magnetosphere surrounding the neutron 
star, there is still a region in the disk where magnetic field stress 
plays an important dynamical role. We construct magnetic slim disk models
appropriate for neutron stars in low-mass X-ray binaries (LMXBs) which
incorporate the effects of both magnetic fields and 
general relativity (GR). The magnetic field --- disk
interaction is treated in a phenomenological manner, allowing for
both closed and open field configurations.
We show that even for surface magnetic
fields as weak as $10^7-10^8$ G, the sonic point of the accretion flow
can be significantly modified from the pure GR value (near $r_{\rm
GR}=6GM/c^2$ for slowly-rotating neutron stars). We derive an analytical
expression for the sonic radius
in the limit of small disk viscosity and pressure.
We show that the sonic radius mainly depends on 
the stellar surface field strength $B_0$ and mass accretion rate $\dot M$
through the ratio $b^2\propto \beta B_0^2/\dot M$, 
where $\beta\simeq |B_\phi/B_z|$ measures the azimuthal pitch angle 
of the magnetic field threading the disk. 
The sonic radius thus obtained 
approaches the usual Alfven radius for high $b^2$ (for which 
a genuine magnetosphere is expected to form), 
and asymptotes to $6GM/c^2$ as $b^2\rightarrow 0$.
We therefore suggest that for neutron stars in LMXBs, 
the distinction between the disk sonic radius and
the magnetosphere radius may not exist; there is only one
``generalized'' sonic radius which is determined by both the GR effect
and the magnetic effect.

We apply our theoretical results to 
the kHz quasi-periodic oscillations (QPOs) observed in the X-ray
fluxes of LMXBs. If these QPOs are associated with the orbital frequency 
at the inner radius of the disk, 
then the QPO frequencies and their correlation with 
mass accretion rate can provide useful diagnostics on the 
(highly uncertain) nature of the magnetic field -- disk interactions. 
In particular, a tight upper limit to the surface 
magnetic field $B_0$ can be obtained, i.e., $B_0\lo 3\times 10^7
(\dot M_{17}/\beta)^{1/2}$ G, where $\dot M_{17}=\dot M/(10^{17}\,{\rm
g\,s}^{-1})$, in order to produce kHz orbital frequency at the sonic 
radius. Current observational data may suggest
that the magnetic fields in LMXBs have complex topology.

\end{abstract}


\section{Introduction}

The inner region of disk accretion onto neutron stars may
be characterized by two unique radii:
(i) The marginally stable orbit due to general 
gravity (GR). For nonrotating neutron stars this is 
located at 
\be r_{\rm GR}={6GM\over c^2}=12.4\,M_{1.4}~{\rm km},
\ee
where $M$ is the neutron star mass, and $M_{1.4}=M/(1.4\,M_\odot)$.
For finite rotation rates, $r_{\rm GR}$ is somewhat smaller.
The flow behavior near $r_{\rm GR}$ has been 
subjected to numerous studies, especially in the context of
black hole accretion disks (e.g., Muchotrzeb \& Paczy\'nski 1982; 
Matsumoto et al.~1984; Abramowicz et al.~1988; Narayan et al.~1997; Chen et
al.~1997): Close to $r_{\rm GR}$ the inward radial velocity of the
accreting gas increases steeply with decreasing radius and becomes
supersonic. The existence of such marginally stable orbit for 
neutron star is predicated on the fact that neutron star models 
constructed using different nuclear equations of state generally give
a stellar radius less than $r_{\rm GR}$ (Arnett \& Bowers 1977;
Klu\'zniak \& Wagoner 1985). 
(ii) The magnetospheric radius, $r_{\rm m}$, 
below which magnetic stress dominates disk plasma stress. 
While the precise value of $r_{\rm m}$ depends on the (rather
uncertain) details of the magnetic field -- disk interactions,
it is estimated to close to or slightly less than (by a factor of a
few) the spherical Alfven radius, i.e., 
\be
r_{\rm m}\simeq \eta\,r_{\rm A}\simeq \eta R\left({B_0^2R^3\over \dot
M\sqrt{GMR}}\right)^{2/7}
=18\,\eta\, R_{10}^{12/7}M_{1.4}^{-1/7}B_8^{4/7}\dot M_{17}^{-2/7}~({\rm km}),
\label{alfven}\ee
with $\eta\lo 1$ (e.g., Pringle \& Rees 1972; Lamb, Pethick \& Pines 1973;
Ghosh \& Lamb 1979; Arons 1987), 
where we have scaled various quantities to values appropriate for
neutron stars in low-mass X-ray binaries (LMXBs):
$R=10\,R_{10}$ km is the neutron star radius, 
$B_0=10^8B_8$ G is the dipolar surface field strength, and
$\dot M=(10^{17}{\rm g~s}^{-1})\dot M_{17}$ is the mass accretion rate
(The Eddington accretion rate is about $10^{18}$ g~s$^{-1}$). 
For highly magnetized neutron stars (such as X-ray pulsars, typically 
having $B\go 10^{12}$ G), $r_{\rm m}$ is much greater than $r_{\rm GR}$ and
the stellar radius, the 
disk is therefore truncated near $r_{\rm m}$, within which
the disk plasma becomes tied to the closed field lines
and is funneled onto the magnetic poles of the star, although 
some plasma may continue to fall in
the equatorial plane as a result of interchange instabilities
(Spruit \& Taam 1990; see also Arons \& Lea 1980).
For weakly magnetized neutron stars, such as those expected in LMXBs,
$r_{\rm m}$ and $r_{\rm GR}$ are comparable, and the plasma
may not climb onto the field lines before reaching the stellar surface.   
A question therefore arises as to how the magnetic field affects the
the dynamics of the inner disk and changes the sonic point.
In this paper, we present an unified (albeit phenomenological)
treatment of neutron star accretion disks under the combined 
influences of magnetic fields and strong gravity.

Our study is motivated by the recent observations 
using the Rossi X-ray Timing Explorer (RXTE) (Bradt, Rothschild \& Swank 1993)
which revealed kilo-Hertz quasi-periodic oscillations (QPOs) in the X-ray
fluxes of at least thirteen LMXBs (see Van der Klis 1997 for a review).
These kHz QPOs are characterized by their high levels of
coherence (with $\nu/\Delta\nu$ up to $100$), large rms amplitudes (up
to $20\%$), and wide span of frequencies ($500-1200$ Hz) which, in
most cases, are strongly correlated with the 
X-ray fluxes. In several sources, the X-ray power spectra show twin kHz 
peaks moving up and down in frequency together, with the separation 
frequency roughly constant. Moreover, in five atoll sources
single QPOs (with a much higher level of coherence)
have been seen during one or more X-ray bursts, with frequencies
equal to the frequency differences between the two peaks or twice that. 
This is a strong indication of beat phenomena (Strohmayer et al.~1996). 
While the origin of these QPOs is uncertain,
it is clear that the action must take place close to the neutron star,
either in the accreting atmosphere (Klein et al.~1996)
or in the inner disk (Strohmayer et al.~1996; Miller, Lamb and Psaltis 1996).
A generic beat-frequency model assumes that 
the QPO with the higher frequency is associated
with the Kepler motion at some preferred orbital radius
around the neutron star, while the lower-frequency QPO results
from the beat between the Kepler frequency and the neutron star 
spin frequency. It has been suggested that this preferred radius
is the magnetosphere radius (Strohmayer et al.~1996) or the sonic
radius of the disk accretion flow (Miller et al.~1996).

In this paper, we are not concerned with the actual mechanisms 
by which kHz QPOs in the X-ray fluxes of LMXBs may be produced
(See Miller et al.~1996 and Van der Klis 1997 for extensive discussion
on various possibilities). Rather, our main purpose is to 
understand what physical effects determine the characteristics
of the inner accretion disks in LMXBs. 
In the sonic-point model, Miller et al.~(1996) suggest that 
some accreting gas can penetrate inside the magnetosphere, whose boundary 
is located at a larger radius than the sonic radius. For unknown 
reasons, they assume that these gases are unaffected by the magnetic 
field once they are inside the magnetosphere and remains
in a Keplerian disk. They further suggest that the variation of QPO frequency
results from the change in radiative forces on the accretion disk.  
We note, however, that the effect of
radiative forces on the disk fluid (rather than test particle orbiting the
central star), 
is far from clear.
Calculating particle trajectories without solving for the global disk
structure (M.~C. Miller 1997, private communication) is inadequate for
determining the magnitude of the radiative forces.
While the radiative forces may be important for high-luminosity
Z-sources, their effects on the disk dynamics are expected to be 
be small for low-luminosity systems ($L$ less than $10\%$ of 
$L_{\rm Edd}$). On the other hand, it is well known 
that millisecond pulsars have magnetic fields in the range
of $10^7-10^9$ G, and one expect that neutron stars in LMXBs to have the 
similar range of field strengths. While the magnetic field
may not be strong enough to induce a corotating magnetosphere
outside the neutron star, it can nevertheless influence the dynamics of
the inner disk flow by transporting away angular momentum from the disk. 

Ideally, to properly assess the dynamical effect of magnetic fields
on the accretion disks, one needs to solve for both the fluids and
the fields self-consistently. This is a difficult task if not impossible. 
Despite many decades of theoretical studies
(e.g., Pringle \& Rees 1972; Lamb et al.~1973;
Ghosh \& Lamb 1979; Aly 1985,~1991; 
Arons 1987; Spruit \& Taam 1990; Sturrock 1991; Shu et al.~1994; 
Lovelace et al.~1987,~1995; Stone \& Norman 1994;
Miller \& Stone 1997), 
there remain considerable uncertainties
on the nature of the stellar magnetic field -- disk interactions.
Particularly outstanding are the issues related to the
efficiency of magnetic field dissipation in and outside the disk
and whether the stellar field
threads the disk in a closed configuration or it becomes open due to
differential shearing between the star and the disk.  
It seems unlikely that some of these issues can be
resolved on purely theoretical grounds.
In this paper, we shall not attempt a self-consistent 
magnetohydrodynamics (MHD) calculations.
Rather, {\it we shall adopt a phenomenological approach} and
consider rather general field configurations. We believe that such 
an approach is useful in bridging the gap between full MHD theories 
and observations. Indeed, as we show in this paper,
if the observed kHz QPOs are associated with 
the sonic point Kepler frequency, then various systematics of kHz QPOs 
should provide useful constraints on the nature of
magnetic field -- disk interactions as well as on the magnetic field 
structure in LMXBs.  

In \S 2 we introduce a model of magnetic slim accretion disk. 
Numerical solutions are presented in \S 3.
Because of various uncertainties in disk parameters, we shall
focus on the simplest models, 
leaving more complete exploration to future studies. 
However, in \S 4 we introduce the notion of
``generalized marginally stable orbit''
including both the GR and magnetic effects. 
We derive an analytical expression for the sonic 
radius, which shows that the sonic point depends mainly on
two parameters characterizing the disk magnetic field (in addition to
the neutron star mass).  
We show how different modes of magnetic field -- disk
interactions can lead to different sonic-point orbital frequencies
and their scalings with the field strength and mass accretion rate.  
Section 5 concerns the equilibrium spin periods of neutron stars
in our slim magnetic disk model.   
Some applications to kHz QPOs in LMXBs are discussed in \S 6,
where we show that our phenomenological approach can be used to learn about 
the physics of magnetic field -- disk interactions.

Unless otherwise noted, we use geometrized units in which the speed of 
light and Newton's gravitation constant are unity.

\section{Slim Magnetic Accretion Disks}

We now consider geometrically thin axisymmetric accretion disk in
steady state, taking into account of the transonic nature of the flow,
and the deviation from Keplerian motion in the inner region of the
disk. Our models generalize the usual ``slim disks'' around
black holes (e.g., Muchotrzeb \& Paczy\'nski 1982; 
Matsumoto et al.~1984; Abramowicz et
al.~1988; Narayan et al.~1997; Chen et al.~1997)
by including the effect of magnetic fields. 
GR effects are included in our purely Newtonian treatment by using the
pseudo-Newtonian potential introduced by Paczy\'nski \& Wiita (1980)
\be
\Phi=-{M\over r-2M}.
\ee
This potential correctly reproduces the marginally stable orbit
\footnote{With small spin parameter $\bar a\equiv J/M^2<<1$
(where $J$ is the spin angular momentum), we have
$r_{\rm GR}/M\simeq 6(1-0.544\bar a)$, and $M\Omega_K(r_{\rm GR})
\simeq 6^{-3/2}(1+0.748\bar a)$. For a spin frequency of $300$ Hz 
(Strohmayer et al.~1996), 
this amounts to a correction of $7\%$ to $r_{\rm GR}$ and
$10\%$ to $\Omega_K$. These corrections are neglected in this paper.}
located at $r_{\rm GR}=6M$, and is adequate for this initial exploration,
considering the much greater uncertainties in the magnetic field -- disk 
interactions. The self-gravity of the disk is neglected. 

We assume that the accreting material is confined to a thin disk, 
and we do not formally introduce a magnetosphere in our model.
As discussed in \S 1, when the field strength is sufficiently high (as in
X-ray pulsars, which typically have $B\go 10^{12}$ G), 
there is no question that a corotating magnetosphere 
exists outside the neutron star surface, located near $r_m$ (see
Eq.~[\ref{alfven}]; note that, theoretically, the magnetosphere radius 
is not known to within a factor of two, nor is it clear what the width
of the transition zone is).
In such a high-field regime, we shall find that the sonic point 
as obtained from our model is approximately equal
to the usual Alfven radius. Although in our model the flow
continues to be confined in the disk plane even inside the
magnetosphere, in reality it may well behave 
differently (e.g., the plasma 
may ``jump'' onto the field lines and get funneled onto the magnetic poles).
Thus for high magnetic systems, the supersonic portion of our flow 
(inside the sonic point) may not be realistic. 
However, for low magnetic systems (such as LMXBs), which is the main focus of 
this paper, there needs not be a genuine magnetosphere to truncate the disk
flow, but the magnetic forces can still shift the sonic point to a radius
larger than $6M$. In such low-field regimes, we expect our global flow 
solutions to have a wider validity.

\subsection{Basic Equations}

The mass continuity equation takes the form
\be
\dot M=-2\pi r\Sigma u,
\label{mdot}\ee
where 
$u$ is the radial velocity of the flow ($u<0$ for
accretion), and $\Sigma=\int\!dz\,\rho\simeq 2H\rho$ is the
surface density of the disk. The disk half-thickness
\footnote{We neglect the magnetic field effect on the disk thickness. 
In reality, the disk can be compressed or flattened depending upon the
field configuration (see Wang et al.~1990 and Stone \& Norman 1994).
However, the subsequent analyses in this paper do not depend upon having
an explicit expression for $H$ since only height-integrated equations
are used.} is given by 
$H=c_s/\Omega_K$, where $c_s=(p/\rho)^{1/2}$ is the 
isothermal sound speed, and $\Omega_K$ is the Keplerian angular
velocity (for the pseudo-Newtonian potential):
\be
\Omega_K=\left({M\over r^3}\right)^{1/2}{r\over r-2M}.
\label{omegak}\ee
The radial momentum equation reads
\be
u{d\,u\over dr}=-{1\over \Sigma}{d\,P\over dr}+(\Omega^2-\Omega_K^2)r
+{B_zB_r\over 2\pi\Sigma}\biggl |_{z=H},
\label{uup}
\ee
where $P=\int\!dz\,p$ is the integrated disk pressure,  
$\Omega$ is the angular velocity.
The last term in Eq.~(\ref{uup}) represents the dominant 
radial magnetic force,
obtained by integrating over height the force per unit volume
$\partial(B_zB_r/4\pi)/\partial z$ and dividing by $\Sigma$.
Note that in Eq.~(\ref{uup}), $B_r|_{z=H}$ is evaluated at the
upper disk plane, and $B_r|_{z=-H}=-B_r|_{z=H}$.
In deriving the magnetic force, we have also neglected 
the $rr$-component of the magnetic stress. 

The angular momentum equation reads
\be
u {d\,l\over dr}={r\over\Sigma}
\left[{1\over r^2}{d\over dr}
\left(r^3\Sigma\nu{d\Omega\over dr}\right)+{B_zB_\phi\over 2\pi}\biggl
|_{z=H}\right],
\label{dl}\ee
where the second term on the right-hand-side is the magnetic torque per unit
mass, obtained by integrating over height the torque per unit volume
$r B_z(\partial B_\phi/\partial z)/4\pi$ and dividing by $\Sigma$.
Equation (\ref{dl}) can be integrated in $r$ to give the conservation equation
for angular momentum:
\be
\dot M l_0=\dot M l+2\pi \nu r^3\Sigma{d\Omega\over dr} + \dot M N_B(r),
\label{l0}
\ee
where $l_0$ is the integration constant, and
\be
\dot M N_B(r)=-\int_r^\infty\!dr\,r^2 B_zB_\phi\biggl|_{z=H}.
\label{N_B}\ee
The three terms on the right-hand-side of Eq.~(\ref{l0})
correspond to advective angular momentum transport, viscous
torque, and magnetic torque due the threading field lines from $r$ to
$\infty$, respectively. 
The constant $l_0$ is the eigenvalue of the problem; it should
be determined by requiring the flow to be regular at the sonic point.
We shall adopt the standard $\alpha-$prescription 
for the disk kinematic viscosity, i.e., $\nu=\alpha H c_s$ 
(Shakura \& Sunyaev 1973).

The energy equation describing the thermal state of the flow
can be written in the form:
\be
\Sigma T u {d S\over dr}=\dot E_{\rm visc}+\dot E_{\rm Joule}
-2F_z.
\ee
Here $S$ is the specific entropy (per unit mass),
$\dot E_{\rm visc}$ is the viscous heating rate per unit area. With the
$\alpha$-prescription for the disk viscosity, we have
\be
\dot E_{\rm visc}=2H{1\over \rho\nu}\left(\rho\nu r{d\Omega\over dr}\right)^2
=\nu\Sigma\left(r{d\Omega\over dr}\right)^2.
\ee
The vertical (optically thick) radiative transport flux is
\be
F_z=-{c\over 3\rho\kappa}{d\over dz}(aT^4)\simeq
{caT^4\over\kappa\Sigma},
\ee
where $\kappa$ 
is the opacity and $aT^4$ is the radiation energy density.
The Joule heating rate $\dot E_{\rm Joule}$ depends on
the field dissipation in the disk, and its specific form depends on
our ansatz for the magnetic field (\S2.2).

Finally we need equations of state. For the inner disk region
of interest in this paper, radiation pressure dominates
over gas pressure. Thus we have $p=aT^4/3$ and $P=2H a T^4/3$.
Also, the opacity is dominated by Thomson scattering, $\kappa=0.4$ cm$^2/$g.

\subsection{Ansatz for the Magnetic Field}

The equations above can be applied to general axisymmetric
magnetic field -- disk configurations, as long as mass loss from 
possible disk wind is negligible, and the accreting material is confined
to a thin disk plane. We now specify our ansatz for the magnetic
fields. The vertical field component is assumed
to take the form
\be
B_z=B_0\left({R\over r}\right)^n.
\label{bz}\ee
We shall mostly focus on the $n=3$ case, corresponding to a central
stellar dipole field threading the disk (e.g., 
Ghosh \& Lamb 1979; K\"onigl 1991; Yi 1995; Wang 1995), although we will also
consider more general values of $n$, as in the cases when 
high-order multipoles are important (Arons 1993) or when field lines become 
open due differential shearing between the disk and the star 
(Aly 1985,~1991; Sturrock 1991; 
Newman et al.~1992; Lynden-Bell \& Boily 1994; Lovelace et al.~1995).
In reality, the power-law relation in Eq.~(\ref{bz}) is most likely to
be valid only for a small range of $r$, but as we shall see
in \S 4, the sonic point is mainly determined by the local
behavior of the magnetic field.

For the azimuthal component of the magnetic field, we consider two
possibilities: 

(i) If the stellar magnetic field threads the 
accretion disk in a closed configurations
(e.g., Ghosh \& Lamb 1979), then $B_\phi$ is governed by
$\partial B_\phi/\partial t=B_z\partial(r\Omega)/\partial z
-B_\phi/\tau$ (where $\tau$ is the field dissipation time).
In steady-state, this gives $B_\phi\sim \tau
(\Omega_s-\Omega)(r/H)B_z$, where 
$\Omega_s$ is the rotation frequency of the star.
We define a dimensionless parameter $\beta$ such that
\be
B_\phi\biggl|_{z=H}=\beta\left({\Omega_s-\Omega\over\Omega_K}\right)B_z,
\label{bphi}\ee
where $\Omega_K$ is the Kepler frequency and $\Omega(\simeq \Omega_K)$ is
the disk orbital frequency.
Various (uncertain) estimates for the field dissipation timescale in the 
magnetically threaded disk configurations have been summarized in Wang (1995). 

(ii) If the magnetic field becomes open (e.g., Lovelace et al.~1995), 
we assume
\be
B_\phi|_{z=H}=-\beta B_z, 
\ee
where $\beta\sim 1$ specifies the
maximum twist angle of any field line connecting the star and the
disk. Clearly, Eq.~(\ref{bphi}) encompasses the second possibility 
if we set $\Omega_s=0$.
However, we note that the physical meaning of $l_0$ in these two cases are 
rather different: For closed field
configurations (i), $\dot M l_0$ measures the total torque
on the star, while for the (partially) open field configurations (ii),
$\dot M l_0$ also include the angular momentum carried away from the
disk by the magnetic fields of disk outflow. 
In both cases, $\dot M l_0$ is the total angular momentum transported away 
from the disk per unit time.

Similar to Eq.~(\ref{bphi}), our ansatz for the radial component $B_r$ 
of the disk magnetic field is 
\be
B_r\biggl |_{z=H}=\beta_r\left({-u\over\Omega_K r}\right)B_z.
\ee
We expect $\beta_r$ to be of the same order of magnitude as $\beta$.

With the particular ansatz for the magnetic fields given by 
Eq.~(\ref{bphi}),
the Joule heating rate $\dot E_{\rm Joule}$ can be calculated as
\be
\dot E_{\rm Joule}=\int\!dz\,{1\over 4\pi}B_zB_\phi
{\partial\over\partial z}(r\Omega)
={1\over 2\pi}\beta r B_z^2{(\Omega_s-\Omega)^2\over\Omega_K}.
\ee
The dissipation due to the current associated with $B_r$ is much smaller
and has been neglected.

\section{Numerical Solutions}

It is useful to define dimensionless field strengths $b$ and $b_r$ via
\begin{eqnarray}
b^2 &=& \beta{B_0^2R^3\over\dot Ml_R}
=7.317\,\beta R_{10}^{5/2}M_{1.4}^{-1/2}{B_8^2\over\dot M_{17}},\label{defb2}\\
b_r^2 &=& \beta_r{B_0^2R^3\over\dot Ml_R}
=7.317\,\beta_r R_{10}^{5/2}M_{1.4}^{-1/2}{B_8^2\over\dot M_{17}},
\end{eqnarray}
where $l_R\equiv (MR)^{1/2}$. Roughly speaking, $b^2$ is the ratio 
of the total magnetic torque and the characteristic accretion torque
on the neutron star. 
Comparing with Eq.~(\ref{alfven}) we see that for dipole magnetic fields,
$r_A/R\sim (b^2/\beta)^{2/7}\sim b^{4/7}$.
In all our calculations, we choose $b_r=b$; The flow
structure and the sonic radius are rather 
insensitive to the value of $b_r$.
The radial force equation and the continuity equation 
can be cast in the form which reveals the existence of a sonic point:
\be
{u^2-a_s^2\over u}\left({du\over dr}\right)={a_s^2\over r}+{l^2\over
r^3}-\Omega_K^2 r +b_r^2{u^2l_R\over r^3\Omega_K}\left({R\over
r}\right)^{2n-3},
\label{up}\ee
where $a_s^2\equiv (dP/d\Sigma)_{\rm flow}$. The sonic point (where $|u|=a_s$)
is a critical point of the differential equation.
The other equations can also be rewritten in the forms convenient for
numerical integration.
The angular momentum equation is
\be
l_0=l-{\alpha r^2 c_s^2\over u\Omega_K}{d\Omega\over dr}
+N_B,
\label{l0_2}\ee
with
\be
{dN_B\over dr}=-b^2{l_R\over r}\left({R\over r}\right)^{2n-3}
{\Omega-\Omega_s\over\Omega_K}.
\label{dnb}\ee
The energy equation is
\be
{\Sigma Tu\over\dot M}{dS\over dr}=-{\alpha c_s^2r
\over 2\pi u\Omega_K}\left({d\Omega\over dr}\right)^2
+{1\over 2\pi}b^2{l_R\over r^2\Omega_K}
\left({R\over r}\right)^3(\Omega_s-\Omega)^2
-{3c c_s\Omega_K\over\kappa\dot M}.
\label{energy_2}
\ee

The eigenvalue $l_0$ is adjusted so that the solution is regular at the
sonic point. Equations (\ref{up})-(\ref{energy_2}) 
are integrated inward from an outer radius (far from 
the sonic point) where the disk is approximately Keplerian. Note that 
in this outer Keplerian region, Eq.~(\ref{dnb}) can be integrated to give
\be
N_B(r)={b^2\,l_R\over 2n-3}\left({R\over r}\right)^{2n-3}\left\{
1-{(4n-6)\Omega_s\over (4n-9)(M/r^3)^{1/2}}
\left[1-{(8n-18)M\over (4n-7)r}\right]\right\},
\label{nbb}\ee
and the angular momentum equation yields
\be
u(r)=-{3\alpha r c_s^2\over 2(l_K-l_0+N_B)}
\left({1-2M/3r\over 1-2M/r}\right),
\ee
where $l_K=r^2\Omega_K$. The sound speed $c_s$ can be obtained 
from the energy equation (\ref{energy_2}). Since the radial 
velocity is small at large radius, the entropy 
advection term can be neglected. 
Substituting $\alpha c_s^2/u$ from Eq.~(\ref{l0_2}) into
Eq.~(\ref{energy_2}), we find
\be
c_s(r)={\kappa\dot M\Omega_K\over 4\pi c}\left[
\left(1-{l_0\over l_K}+{N_B\over l_K}\right)
\left({1-2M/3r\over 1-2M/r}\right)
+{2\over 3}\,b^2\,{l_R\over l_K}\left({R\over r}\right)^3
\left(1-{\Omega_s\over\Omega_K}\right)^2\right].
\label{cs}\ee

Equations (\ref{up})-(\ref{energy_2}) turn out to be 
a rather stiff set of equations. We have opted to adopt a further 
simplification by assuming the disk is isothermal. 
We estimate the range of $c_s$ using the thin disk 
expression (\ref{cs}) evaluated near the sonic radius.
The physical rationale behind this approximation is that 
the transition from Keplerian disk to supersonic flow is
very sharp, and we do not expect the sound speed 
to change significantly in this transition region (near the sonic point). 
As we shall see in \S 4, the sonic radius is insensitive to 
the thermal state of the disk when the sound speed
is small. We have not studied the general dependence of the sonic radius on 
the thermal state of the disk. However, considering the very large
uncertainties in the disk magnetic fields, the 
isothermal approximation should be adequate for 
use in the first step in our investigation.

Figure 1 depicts two examples of transonic accretion flows, 
with closed dipolar stellar fields threading the disks ($n=3$). 
We choose a standard set of parameters:
$\alpha=0.1$, $c_s=0.01$, $R/M=5$ (a typical value for 
a $M=1.4\,M_\odot$, $R=10$ km neutron star), 
and $\Omega_s=0.013\,M^{-1}$
(corresponding to stellar spin frequency of $300\,M_{1.4}^{-1}$ Hz).
As expected, the magnetic fields slow down the tangential flow velocity,
and, together with the strong relativistic gravity, 
make the radial velocity supersonic at small radii. 
For small $b^2$, the deviation of $\Omega$ from $\Omega_K$ is small;
for larger $b^2$, the sonic radius $r_s$ larger, and
$\Omega$ gradually approaches $\Omega_s$ inside the sonic point. 
In these examples, the sonic points are located at
$r_s/M=7.43$ (for $b^2=1$) and $10.95$ (for $b^2=10$),
the corresponding eigenvalues ($l_0$) are $3.850M$ and $4.153M$,
respectively.

In Figure 2 we show the sonic radius $r_s$ and the corresponding 
specific angular momentum $l_0$ as a function of $b^2$
for several different values of $\alpha$ and $c_s$. 
We assume $R/M=5$ and $\Omega_s=0.013/M$ for these models. 
The following trends have been found: For a given $b^2$, 
a larger $\alpha$ tends to make $r_s$ larger (i.e., 
viscosity tends to ``destablize'' the disk), while a larger $c_s$ tends to
make $r_s$ smaller (i.e., pressure ``stablizes'' the disk).  
However, we emphasize that dependences of $r_s$ on these 
disk parameters ($\alpha, c_s$) are rather weak. Moreover, as 
$c_s$ decreases, the sonic $r_s$ converges to a value independent of
$\alpha$ and $c_s$ --- The reason for this convergence 
will become clear in \S 4.

\section{Analytical Approximation to the Sonic Radius}

As our numerical results in \S 3 indicate, in the limit of 
small disk pressure and viscosity, the sonic radius approaches 
a value independent of the disk parameters ($\alpha,~c_s$
and the equation of state). This {\it asymptotic sonic
radius} can be derived analytically using a simple mechanical model:  
Consider the equation of motion of a test mass around a
neutron star
\be
{d^2 r\over dt^2}={l^2\over r^3}-\Omega_K^2 r,~~~~~
l_0=l+N_B(r).
\label{testeq}\ee
Imagine that the test mass is ``attached'' to a magnetic field line
so that its orbital angular momentum $l$ is not conserved by itself. 
The conserved angular momentum $l_0$ includes
the contributions from both the orbital motion and
the angular momentum $N_B$ [cf.~Eq.~(\ref{N_B})]
carried by the threading magnetic field. 
Equation (\ref{testeq}) can be obtained by setting the pressure and
viscosity to zero in our general slim disk equations (\S 2.1). The radial 
component of the magnetic force has been neglected.
An equilibrium orbit is determined by the condition
\be
l_0-N_B=\Omega_K^2 r=l_K(r).
\ee
Deviation $\Delta r$ from the equilibrium
is governed by the perturbation equation of the form
\be
{d^2{\Delta r}\over dt^2} +\kappa_{\rm eff}^2\Delta r=0,
\ee 
where the the effective epicyclic frequency $\kappa_{\rm eff}$ is given by
\be
\kappa_{\rm eff}^2={2\Omega_K\over r}{d\over dr}(l_K+N_B)=
{M(r-6M)\over r(r-2M)^3}-2\,b^2\,{l_R\Omega_K\over r^2}
\left({R\over r}\right)^{2n-3}\!\left(1-{\Omega_s\over\Omega_K}\right),
\label{kappa2}\ee
(Recall that $l_R=\sqrt{MR}$ and $R$ is the neutron star radius). 
Setting $\kappa_{\rm eff}=0$, we obtain a critical orbit, 
which we dub the {\it ``generalized marginal stable orbit''}, 
located at $r=r_{\rm MSO}$. 
Clearly, $r_{\rm MSO}$ depends only on the 
gravitational potential and the local 
magnetic torque $(dN_B/dr)=r^2B_zB_\phi/\dot M$.
The magnetic field enters only through the dimensionless ratio
$b$. The corresponding constant eigenvalue 
$l_0=l_K(r_{\rm MSO})+N_B(r_{\rm MSO})$, however, depends on the global field
structure. 

In Figure 2 we plot $r_{\rm MSO}$ and $l_0$ against $b^2$ for
$\Omega_s=0.013/M$ (corresponding to spin frequency of $300\,M_{1.4}^{-1}$
Hz) and $R/M=5$. We see that $r_{\rm MSO}$ is
the upper limit to the numerically determined $r_s$, i.e., 
{\it $r_{\rm MSO}$ is the asymptotic sonic radius as the disk 
viscosity and pressure diminish}. 

It is of interest to consider two limiting cases:
(i) In the Newtonian limit (neglecting the GR effect), or equivalently
when $b^2$ is large (so that $r_{\rm MSO} >>6M$), we have
\be
{r_{\rm MSO}\over
R}=\left[2\,b^2\left(1-{\Omega_s\over\Omega_K(r_{\rm MSO})}\right)
\right]^{2/(4n-5)}.
\label{largeb}\ee
For $n=3$, this is the standard result for the inner radius of the
Keplerian disk, as determined by $\dot M (d\,l_K/dr)
=-r^2B_zB_\phi\bigl |_{z=H}$ (e.g., Arons 1993; Wang 1995);
(ii) In the limit of small $b^2$ (so that $r_{\rm MSO}$ is
close to $6M$), we find
\be
{r_{\rm MSO}\over M}=6+{16\,b^2\over 3}\left({R\over
M}\right)^{(4n-5)/2}\left[1-{\Omega_s\over\Omega_K(6M)}\right].
\label{smallb}\ee
This gives the correction to the standard general-relativity-induced
MSO located at $r_{\rm GR}=6M$. 

The consideration of the limiting cases clearly indicates
the sonic point (or the generalized MSO) constructed in our slim disk model
includes the essential physics embodied in the determination 
of the usual magnetosphere radius. As we argued at the beginning 
of \S2, for high magnetic systems, a genuine magnetosphere 
should certainly exist outside the neutron star. For low magnetic
systems, however, the accretion flow may well be confined in the disk, 
and the distinction between the sonic point and 
the magnetosphere boundary may not exist.

We emphasize that our analysis given here is phenomenological.
It only takes account of the dynamics of disk under a fixed magnetic field 
configuration, while a full MHD treatment should include perturbations
of both disk fluid and magnetic fields. 
The usefulness of our analytical result lies in the fact that 
$r_{\rm MSO}$ provides a good approximation to the sonic radius;
and the sonic point is induced by both the GR effect 
and the magnetic effect. 
We note that in the presence of disk viscosity and 
magnetic fields, a fluid element continuously falls inward by the viscous 
stress and the magnetic stress, and therefore the concept of 
``marginally stable orbit'' does not strictly apply. 
Nevertheless, we use the term ``generalized MSO'' to refer to
the asymptotic sonic radius as determined by our analytical expressions. 

Figure 3 depicts the orbital frequency
\footnote{To be consistent with the relativistic expression, 
we use $\nu_{\rm MSO}=(M/r^3_{\rm MSO})^{1/2}/(2\pi)$, rather than the 
approximate pseudo-Newtonian Eq.~(\ref{omegak}).}
at the generalized MSO (approximately the sonic radius)
as a function of the mass accretion rate $\dot M$ for
several different combinations of model parameters ($n$ and $\Omega_s$). 
Note that once we specify $n$ and $\Omega_s$ in units of $1/M$,
the numerical value of $r_{\rm MSO}/M$ depends
on the other parameters only through the combination
$b^2(R/M)^{(4n-5)/2}$ (see Eqs.~[\ref{kappa2}]-[\ref{smallb}]).
We have therefore used 
\be
x=M_{1.4}^{2n-2}R_{10}^{-2n}{\dot M_{17}\over \beta\,B_7^2}
={0.07317\over b^2}\left({M_{1.4}\over R_{10}}\right)^{(4n-5)/2}
\label{xlabel}\ee
as the $x$-variable in Fig.~3 (where $B_7$ is the surface field 
$B_0$ in units of $10^7$ G).
For large $\dot M/(\beta B_7^2)$ (or small $b^2$), 
$r_{\rm MSO}$ approaches $r_{\rm GR}=6M$ and $\nu_{\rm MSO}=
\nu_K(r_{\rm MSO})$ approaches $1.57/M_{1.4}$ kHz.
The scaling of $\nu_{\rm MSO}$ with $\dot M$ depends mainly on 
$n$, the index which specifies the shape of the magnetic field lines
(see Eq.~[\ref{bz}]).
Note that the sonic point converges to
the MSO only in the limit of small $c_s$. Thus in reality,
the dependence of the sonic-point Kepler frequency 
$\nu_K(r_s)$ on $\dot M$ may be different
if $c_s$ is not small. For example, $c_s\propto \dot M$ for
radiation-dominated optically-thick disk, and therefore
the dependence of $\nu_K(r_s)$ on $\dot M$ is slightly steeper than
what is shown in Fig.~3. We shall discuss some of the applications 
of Fig.~3 to QPOs in LMXBs in \S6.

\section{Equilibrium Spin Period of Neutron Star}

As discussed before (\S 2.2), for closed field configurations,
the quantity $\dot M l_0$ measures the total torque on the
neutron star due to the accreting matter and the threading magnetic
fields. It is of interest to consider how the equilibrium stellar
rotation rate $\Omega_{s,eq}$, at which $l_0=0$, is determined in
the slim disk model. We shall restrict to the dipole fields ($n=3$) 
in this section. 

First consider the result when the GR effect is neglected.
In this case the Keplerian disk boundary $r_m$ is determined by
the condition $\dot M(d\,l_K/dr)=-r^2B_zB_\phi\bigl|_{z=H}$, 
or $\dot M l_K(r_m)=2 r_m^3\beta B_z^2(r_m)(1-\omega_s)$,
where $\omega_s\equiv \Omega_s/\Omega_K(r_m)$.
We find
\be
{r_m\over R}=\left[2\,b^2\left(1-\omega_s\right)
\right]^{2/7},
\label{ghosh}\ee
(cf.~Eq.~[\ref{largeb}]). The total torque
on the star is given by
\be
N_{\rm tot}=\dot M l_0=\dot M l_K(r_m){7\over 6}
\left[{1-(8/7)\omega_s\over 1-\omega_s}\right].
\label{torquen}\ee
Equilibrium is obtained for $\omega_s=7/8$. Using Eq.~(\ref{ghosh})
we then find $r_m=(b^2/4)^{2/7}R$ and
$\Omega_{s,eq}=({7/8})({M/R^3})^{1/2}({b^2/4})^{-3/7}$, which gives
\be
\nu_{s,eq}=3.44\,M_{1.4}^{1/2}R_{10}^{-3/2}\,b^{-6/7}\,({\rm kHz})
=1.467\,M_{1.4}^{5/7}R_{10}^{-18/7}\dot M_{17}^{3/7}
\left(\beta B_8^2\right)^{-3/7}\,({\rm kHz}).
\label{omegeq}\ee
This is the standard result that the equilibrium spin frequency is
equal to the Keplerian frequency at the Alfven radius.
This result is plotted in Fig.~4. 

In the slim magnetic disk model, the total torque on the neutron star 
$\dot M l_0$ is obtained from the eigenvalue $l_0$. In the asymptotic
regime discussed in \S4, $l_0$ can be determined from 
the analytical expressions: 
$r_s$ is obtained from the condition $\kappa_{\rm eff}^2=0$ 
in Eq.~(\ref{kappa2}),
and then $l_0=l_K(r_s)+N_B(r_s)$ with $N_B$ given by Eq.~(\ref{nbb})
(specialized to $n=3$). 
It is straightforward to show that 
in the limit of $r>>M$, we recover the results given in
Eqs.~(\ref{ghosh}-\ref{omegeq}).
The equilibrium $\Omega_{s,eq}$ is obtained
by requiring $l_0=0$. In Fig.~4 we plot the equilibrium spin frequency 
$\nu_{s,eq}$ as a function of $b^2$. Clearly, for large $b^2$, 
our calculation agrees with the usual nonrelativistic result. 
Relativistic corrections are significant only when $b^2$ is small, 
for which the sonic point lies close to the neutron star. 

We note that the discussion in this section is valid only 
if the magnetic fields are closed throughout the disk.
Only in these cases does $l_0$ measure the net torque on the
neutron star. Thus our results in this section are much more
restrictive compared to the location of the sonic point (which 
depends only on the local field structure) discussed in \S\S 3-4.

\section{Discussion: Applications to QPOs in LMXBs}

We have presented in \S\S 2-4 an unified treatment of
the inner region of accretion flow under the combined influences of
general relativistic gravity and stellar magnetic fields in LMXBs.  
We have shown that even relatively weak magnetic fields
($10^7-10^8$ G) can slow down the orbital motion in the inner disk
by taking away angular momentum from the disk, thereby 
changing the position of the sonic point
significantly (cf.~Figs.~2-3). 

While the mechanisms responsible for the kHz QPOs in
LMXBs are still uncertain, it is tempting 
to associate them with orbital motions at a certain preferred radius. 
If this is the case, then the Keplerian frequency at inner radius 
of the disk, or more precisely the 
sonic radius is certainly a natural choice 
(see Paczy\'nski 1987 for an earlier suggestion on the importance of
the disk sonic point).
One can envisage a number of different mechanisms
that will lead to QPOs at $\nu_K(r_s)\simeq\nu_{\rm MSO}$ 
and the beat frequency
with the stellar spin (Miller et al.~1996; Van der Klis 1997).
The following discussion is based on
the hypothesis that the kHz QPO frequency corresponds to
$\nu_{\rm MSO}$ or the sonic-point Kepler frequency.
We note that although our treatment of the magnetic field effects 
in \S 2-4 is rather general (albeit based on a phenomenological 
prescription), other physical effects might be at work
(such as radiation forces for high-luminosity systems; see \S 1). 
As a result, some of our conclusions below should be considered tentative.

(a) {\it Range of QPO frequencies and constraint on the magnetic field
strength}: 
As emphasized by Van der Klis (1997), similar QPO frequencies
($500-1200$ Hz) have been observed in sources with widely different
average luminosities $\langle L\rangle$
(from a few times $10^{-3}L_{\rm Edd}$ to near $L_{\rm Edd}$, corresponding 
to $\langle\dot M\rangle$ between a few times $10^{15}$ g~s$^{-1}$
to $10^{18}$ g~s$^{-1}$),
while for an individual source $\nu_{\rm QPO}$ often
correlates strongly with the X-ray flux.
This peculiar lack of correlation between the QPO frequency
and $\langle L\rangle$ can be explained in our model,
as long as the star's magnetic field strength correlates with 
its $\langle\dot M\rangle$ in such a way as to leave $\langle\dot
M\rangle/(\beta B_0^2)$ in a certain range (see Fig.~3). 
For example, if we use the model with $n=3$ (dipole field) and 
$\nu_s=300$ Hz, then to produce $\nu_{MSO}$ in the range of
$500-1200$ Hz would require $0.003\lo \dot M_{17}/(\beta B_7^2)\lo 0.1$;
on the other hand, if we use the model with $n=2$ and $\nu_s=0$
(open field configuration), we would require 
$0.05\lo \dot M_{17}/(\beta B_7^2)\lo 0.3$ (we have adopted 
$M_{1.4}=R_{10}=1$ in these examples). 
Indeed, such correlation between the magnetic field strength and
the mean mass accretion rate among LMXBs
has been suggested independently on the basis
of Z and atoll source phenomenology (Hasinger \& Van der Klis 1989),
although the origin for this correlation is still unclear. 
We note, however, that the correlation needs not be very strong,
considering the wide range of other controlling parameters such as
$n$ (the shape of the magnetic field) and $\nu_s$ (stellar rotation)
(see Fig.~3). 

While there is a natural upper limit to $\nu_{\rm MSO}$ (corresponding
to $r_{\rm MSO}\rightarrow 6M$; but see (c) below), the existence of
a (source-dependent) lower limit to the observed QPO frequency needs
an explanation. If we rely on inner disk accretion to explain 
these QPOs (e.g., Van der Klis 1997), then one 
possibility is that for large $r_{\rm MSO}$ (small
$\nu_{\rm MSO}$), the accreting gas can be channeled out of the disk
plane by the magnetic field ---
This must happen for sufficiently small $\dot M$ (or sufficiently 
large $B$, as in the case of accreting X-ray pulsars). 
The precise location where the plasma leaves the disk
depends on the near-zone field structure, and is clearly
source-dependent.

The observed QPO frequencies may already be used to
probe the magnetic fields in LMXBs and 
the nature of the magnetic field -- disk interactions. 
As an example, consider the atoll source 4U 0614+091
which has extremely small luminosity
(see Ford et al.~1997 and references therein): 
To obtain $\nu_{\rm MSO}\sim 1150$ Hz requires $\dot M_{17}/(\beta B_7^2)
\go 0.1$ (This constraint depends somewhat on the field
structure and $\nu_s$; see Fig.~3). At $\dot M_{17}\sim 0.05$, this translates
to $\beta^{1/2}B_7\lo 0.7$ --- a stronger magnetic field would push the sonic
point (or the generalized MSO) to a larger radius. 
We are left with two possibilities: 
(i) If $B_7\go 10$, we would require $\beta\lo 0.01$, i.e.,   
the dissipation of toroidal fields near the disk must be very efficient;
(ii) If $\beta\sim 1$ (as expected for open field configurations),
then we would require $B_7\lo 1$. Indeed, if the kHz QPO sources
represent a fair sample of LMXBs, 
then we might conclude that the magnetic fields in LMXBs are 
systematically weaker than those in millisecond pulsars. 
This may indicate that the magnetic field of a neutron star 
is ``buried'' during the LMXB phase
(e.g., Romani 1990; Urpin \& Geppert 1995; Konar \& Bhattacharya 1997;
Brown \& Bildsten 1998), and later regenerates or re-emerges
as accretion stops. 

(b) {\it Scaling of $\nu_{\rm QPO}$ with $\dot M$}: 
For most sources\footnote{The possible exceptions are
4U~1608-52 (Berger et al.~1997)
and the high-intensity (``banana'') state of 4U~1636-53
(Wijnands et al.~1997), 
for which the power spectrum shows a single peak
with frequency independent of the count rate. However, this may be
an artifact of insensitive search of weak QPOs; see Van der Klis (1997).},
it was found that the kHz QPO frequency strongly 
correlates with the XTE count rate ($2-50$ keV), with power-law index
greater than unity. However, the scaling relation between
the count rate and $\dot M$ is not well established, 
and it has been suggested the flux of the black-body component is a
better indicator of QPO frequency (Ford et al.~1997).
As discussed in \S 3, $\nu_{\rm MSO}$
depends primarily on the magnetic field structure
near the sonic point, particularly on the ``field shape'' 
index $n$. If the scaling of 
$\nu_{\rm QPO}$ with $\dot M$ can be established observationally,
it may be possible to distinguish a closed field configuration from an
open one. For example, if we believe the scaling $\nu_{\rm QPO}\propto
\dot M^\sigma$ with $\sigma> 1$, then we may conclude that $n<2$ (see
Eq.~[\ref{largeb}] and the discussion following Eq.~[\ref{xlabel}]),
which indicates that magnetic fields in LMXBs do not have dipolar shape,
but rather have complex topology (see Arons 1993 for discussion on 
related issues).

(c) {\it The maximum value of $\nu_{\rm QPO}$}: 
It has been suggested (Zhang et al.~1997) based on the narrow range
of the maximal QPO frequencies ($1100-1200$ Hz) in at least six
sources that these maximum frequencies correspond to the Kepler
frequency at $r_{\rm GR}=6M$, which then implies that the neutron star
masses are near $2M_\odot$ (see also Kaaret et al.~1997).
While we agree that this conclusion seems most natural,
we nevertheless add the following cautionary notes:
(i) The inferred large stellar masses may be problematic:
All neutron stars with well-determined masses 
(including a few that certainly had accreted mass, although not
necessarily in the same accretion mode as in LMXBs) 
have masses consistent with being $M\simeq 1.4\,M_\odot$
(e.g., van Kerkwijk et al.~1995). 
In particular, the $5.4$ ms recycled
pulsar B1855+09, which is thought to have gone through a LMXB phase
(Phinney \& Kulkarni 1994),
has a mass $1.50\pm^{0.26}_{0.14}M_\odot$ 
(Kaspi et al.~1994). Moreover,
accretion of $0.6M_\odot$ might have spun up
the neutron stars to near break-up (see Cook et al.~1994 for
calculations of spin-up tracks in the nonmagnetic case), in contrary to
the observed spin rates ($300-500$ Hz). 
(ii) If the maximum $\nu_{\rm QPO}$ is indeed
$\nu_K(6M)$, then the correlation between $\nu_{\rm QPO}$ and $\dot M$
should weaken as $\dot M$ increases, 
and eventually $\nu_{\rm QPO}$ should approaches a constant
independent of $\dot M$ (see Fig.~3).
This has not been observed. Therefore in our
opinion it is premature to identify the maximal $\nu_{\rm QPO}$ with the
Kepler frequency at $6M$. An alternative is that as 
$\nu_{\rm QPO}$ approaches $1100-1200$ Hz, the rms QPO amplitude
decreases --- as the observations have indicated, making it difficult
to detect higher QPO frequencies. 

(d) {\it Horizontal-Branch Oscillations (HBOs) in Z-sources}: 
In several Z-sources (e.g., Sco X-1, GX 5-1 and GX 17+2),
HBOs with frequencies $20-50$ Hz have been detected {\it simultaneously}
with the kHz
QPOs (see Van der Klis 1997). One standard interpretation of HBOs 
is that they are associated
with the beat between the Kepler frequency at the
magnetosphere boundary and the neutron star spin
(Alpar \& Shaham 1985). Since the spin frequencies $\nu_s$ of 
these sources (as determined from the difference in the twin kHz QPO
frequencies) lie around $300$ Hz (see White \& Zhang 1997), 
the putative magnetosphere boundary must be located at a large radius 
where $\nu_K\sim \nu_s\sim 300$ Hz. 
As we have shown in this paper, such a strong magnetic field must 
necessarily push the (generalized) disk sonic
point to a large radius where the Kepler frequency 
drops below the kilo-Hertz range.
(Recall that for low field systems such as LMXBs, the 
distinction between the sonic point and the magnetosphere boundary
probably does not exist, and the two separate radii are replaced by a
single generalized sonic point [see \S\S3-4].)
Therefore if the kHz QPOs 
are associated with the sonic-point Kepler frequency, 
then the magnetospheric beat frequency model for HBOs cannot work, and 
the origin of HBOs must lie elsewhere.
Alternative models for HBOs have been
discussed by Biehle \& Blandford (1993) and Stella \& Vietri (1997).

\acknowledgments

The author thanks Zhiyun Li, Rob Nelson and Brian Vaughan for valuable
discussions and Lars Bildsten for comment. He also thanks
the referee for constructive comments which improved the presentation
of the paper. This research is supported by a Richard Chace Tolman
Fellowship at Caltech, NASA Grant NAG 5-2756, and NSF Grant
AST-9417371. 


\clearpage

\begin{figure}
\plotone{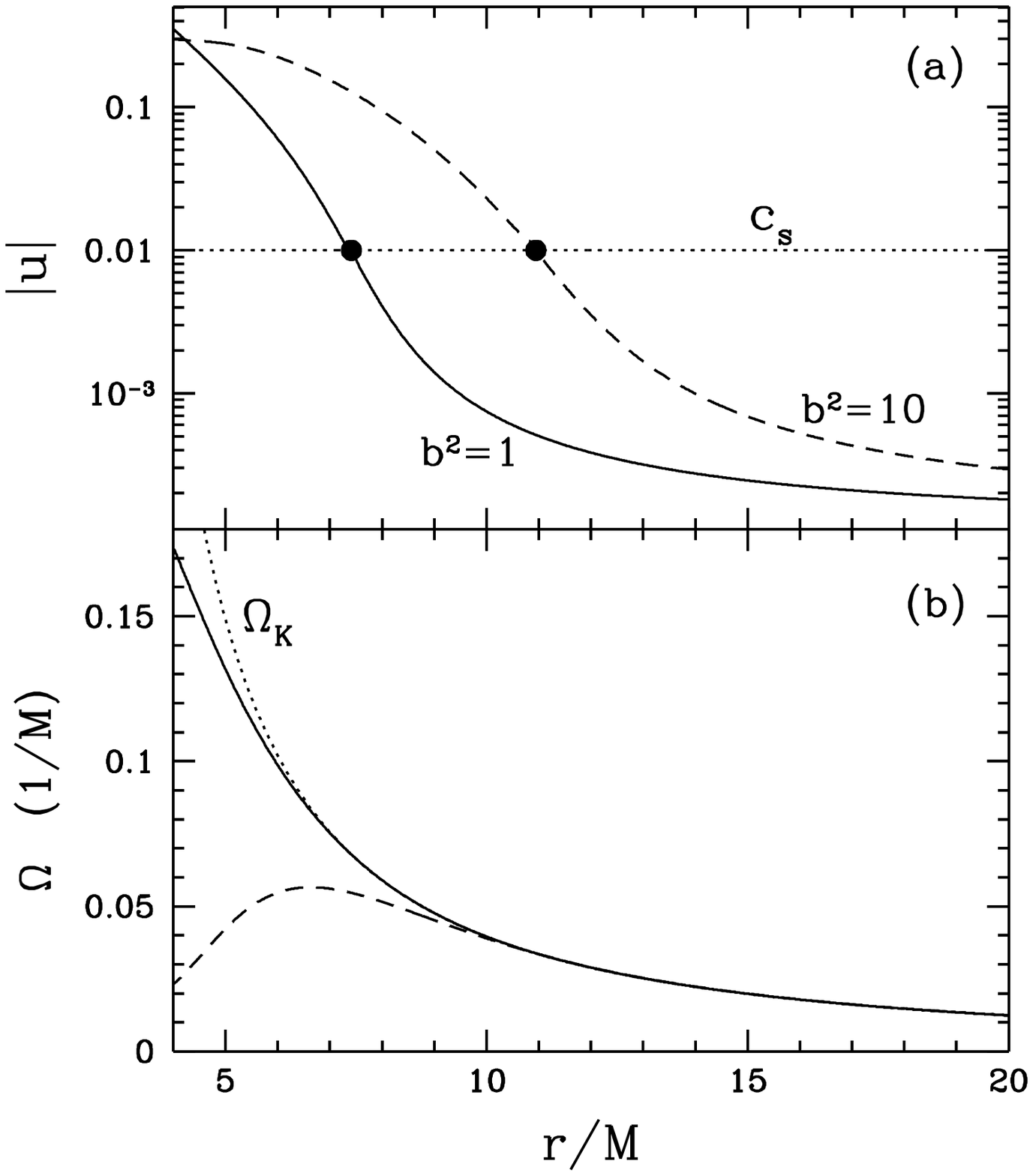}
\caption{
(a) Radial velocity $|u|$ (in units of $c$) 
and (b) angular velocity $\Omega$ (in units
of $c^3/GM$) of transonic accretion flows as a function of disk 
radii. The model parameters are $n=3$ (dipolar field),
$R/M=5$, $\alpha=0.1$, $c_s=0.01$, $\Omega_s=0.013/M$, and $b^2=1$ (solid
curves), $10$ (dashed curves). The sonic points are marked by filled circles.
The dotted lines depict the sound speed (a) and Keplerian angular
velocity (b).
\label{fig1}}
\end{figure}

\begin{figure}
\plotone{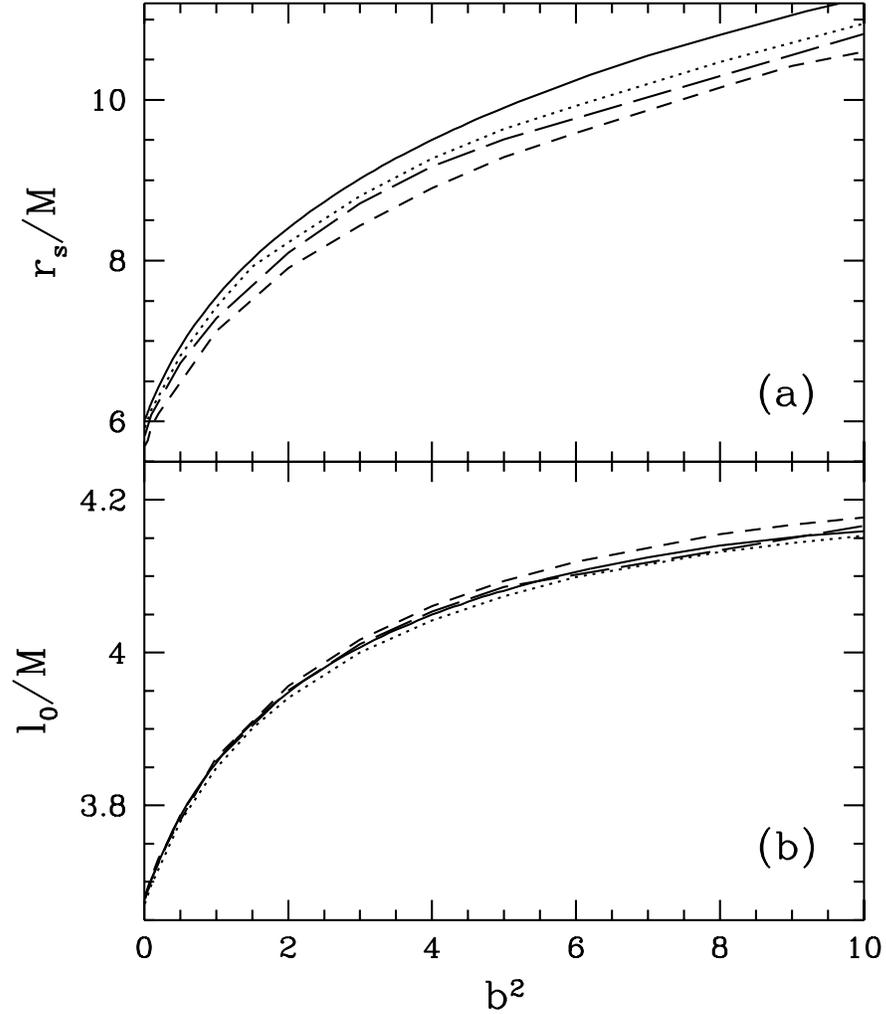}
\caption{
(a) The sonic radius and (b) the constant angular momentum $l_0$
as a function of the dimensionless ratio $b^2$ 
(defined in Eq.~[18])
for three different models:
The dotted lines are for $\alpha=0.1,~c_s=0.01$,
the short-dashed lines for $\alpha=0.02,~c_s=0.01$,
and the long-dashed lines for $\alpha=0.02,~c_s=0.005$.
All models have $n=3$, $R/M=5$ and $\Omega_s=0.013/M$.
The solid lines are the asymptotic analytic solutions
as discussed in \S 4.
\label{fig2}}
\end{figure}

\begin{figure}
\plotone{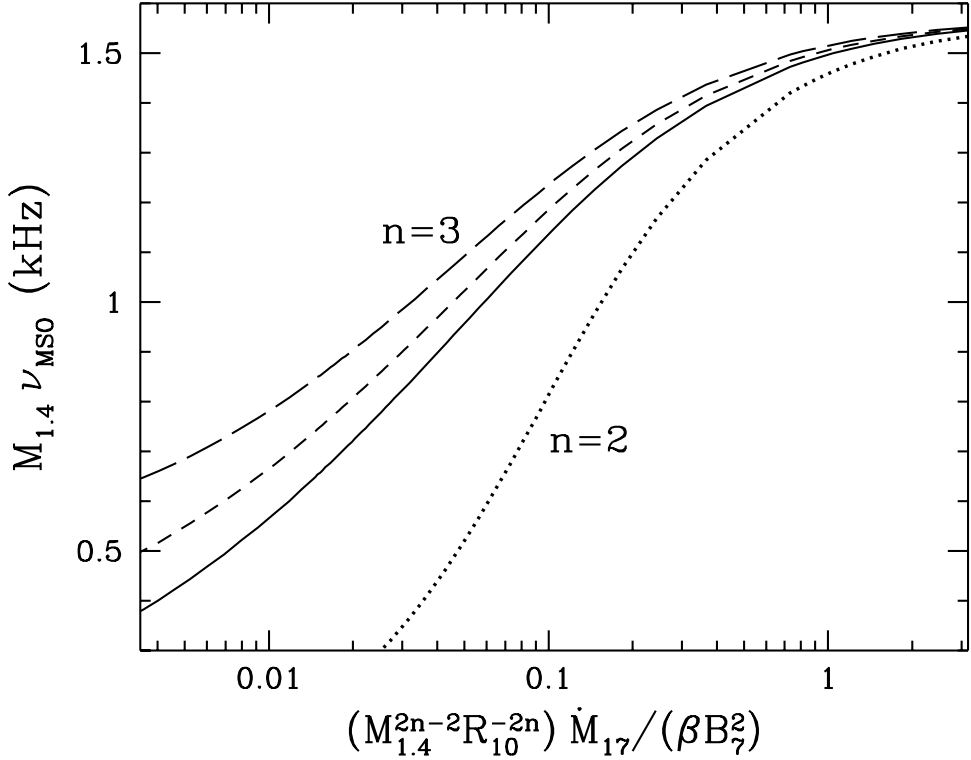}
\caption{
Orbital frequency $\nu_{\rm MSO}=\nu_K(r_{\rm MSO})$ at the generalized MSO 
(which approximates the sonic point)
as a function of mass accretion rate. The results are obtained 
using the analytical expressions given in \S 4. 
The dotted line corresponds
to an open field configuration with $\Omega_s=0$ and $n=2$,
while the other three lines correspond to closed field configurations
with $n=3$, and $\Omega_s=0$ (solid line), $0.013/M$ (dashed line),
and $0.026/M$ (long-dashed line). 
Note that for large $\dot M/(\beta B_7^2)$ (or small $b^2$), 
$r_{\rm MSO}$ approaches $r_{\rm GR}=6M$ and $\nu_{\rm MSO}$
approaches $1.57/M_{1.4}$ kHz.
\label{fig3}}
\end{figure}

\begin{figure}
\plotone{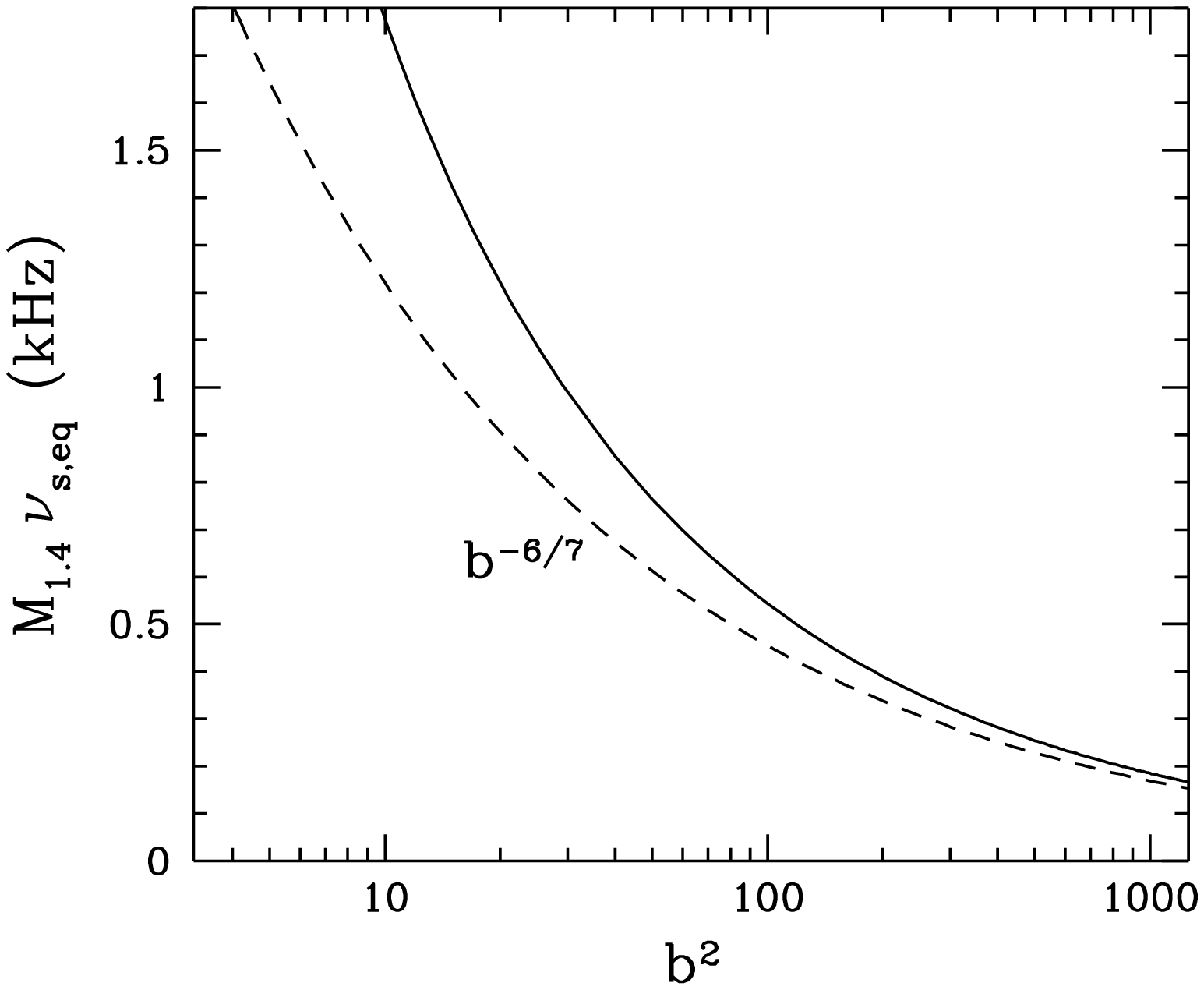}
\caption{
The equilibrium spin frequency $\nu_s$ of the neutron star as a function 
of the dimensionless ratio $b^2$ 
(defined in Eq.~[18]).
The solid line is our result including both the GR and magnetic field
effects, and the dashed line is usual result as given by 
Eq.~(36).
\label{fig4}}
\end{figure}


\begin{thebibliography}{}

\bibitem[]{} 
Abramowicz, M.~A., Czerny, B., Lasota, J.~P., \& Szuszkiewicz,
E. 1988, ApJ, 332, 646.

\bibitem[]{} 
Alpar, A., \& Shaham, J. 1985, Nature, 316, 239.

\bibitem[]{} 
Aly, J.~J. 1985, A\&A, 143, 19.

\bibitem[]{} 
Aly, J.~J. 1991, ApJ, 375, L61.


\bibitem[]{} 
Arnett, W.~D., \& Bowers, R.~L. 1977, ApJS, 33, 415.

\bibitem[]{} 
Arons, J. 1987, in ``The Origin and Evolution of Neutron Stars'' (IAU
Symp.~No.~125), ed. D.~J. Helfand \& J.-H. Huang (D. Reidel Pub.:
Dordrecht).

\bibitem[]{} 
Arons, J. 1993, ApJ, 408, 160.

\bibitem[]{} 
Arons, J., \& Lea, S.~M. 1980, ApJ, 235, 1016.

\bibitem[]{} 
Berger, M., et al. 1996, ApJ, 469, L13. 

\bibitem[]{} 
Biehle, G.~T., \& Blandford, R.~D. 1993, ApJ, 411, 302.

\bibitem[]{} 
Bradt, H.~V., Rothschild, R.~E., \& Swank, J.~H. 1993, A\&AS, 97, 355.

\bibitem[]{} 
Brown, E.~F., \& Bildsten, L. 1998, ApJ, in press (astro-ph/9710261). 

\bibitem[]{} 
Chen, X., Abramowicz, M.~A., \& Lasota, J.-P. 1997, ApJ, 476, 61.

\bibitem[]{} 
Cook, G.~B., Shapiro, S.~L., \& Teukolsky, S.~A. 1994, ApJ, 423, L117.

\bibitem[]{} 
Ford, E.~C., et al. 1997, ApJ, 486, L47 (astro-ph/9706100).

\bibitem[]{} 
Ghosh, P., \& Lamb, F.~K. 1979, ApJ, 232, 259.

\bibitem[]{} 
Hasinger, G., \& Van der Klis, M. 1989, A\&A, 225, 79.

\bibitem[]{} 
Kaaret, P., Ford, E.~C., \& Chen, K. 1997, ApJ, 480, L27.

\bibitem[]{} 
Kaspi, V.~M., Taylor, J.~H., \& Ryba, M.~F. 1994, ApJ, 428, 713.

\bibitem[]{} 
Klein, R.~I., Arons, J., Jernigan, G., \& Hsu, J. 1996, ApJ, 457, L85.

\bibitem[]{} 
Klu\'zniak, W., \& Wagoner, R.~V. 1985, ApJ, 297, 548.

\bibitem[]{} 
Konar, S., \& Bhattacharya, D. 1997, MNRAS, 284, 311.

\bibitem[]{} 
K\"onigl, A. 1991, ApJ, 370, L39.

\bibitem[]{} 
Lamb, F.~K., Pethick, C.~J., \& Pines, D. 1973, ApJ, 184, 271.

\bibitem[]{} 
Lovelace, R.~V.~E., Wang, J.~C.~L., \& Sulkanen, M.~E. 1987, ApJ, 315, 504.

\bibitem[]{} 
Lovelace, R.~V.~E., Romanova, M.~M., \& Bisnovatyi-Kogan, G.~S. 1995,
MNRAS, 275, 244.

\bibitem[]{} 
Lynden-Bell, D., \& Boily, C. 1994, MNRAS, 267, 146.

\bibitem[]{} 
Matsumoto, R., Kato, S., Fukue, J., \& Okazaki, A.~T. 1984, PASJ, 36, 71.


\bibitem[]{} 
Miller, M.~C., Lamb, F.~K., \& Psaltis, D. 1996, ApJ, submitted
(astro-ph/9609157).

\bibitem[]{} 
Miller, K.~A., \& Stone, J.~M. 1997, ApJ, 489, 890.

\bibitem[]{} 
Muchotrzeb, B., \& Paczy\'nski, B. 1982, Acta Astr., 32, 1.

\bibitem[]{} 
Narayan, R., Kato, S., \& Honma, F. 1997, ApJ, 476, 49.

\bibitem[]{} 
Newman, W.~I., Newman, A.~L., \& Lovelace, R.~V.~E. 1992, ApJ, 392, 622.


\bibitem[]{} 
Paczy\'nski, B. 1987, Nature, 327, 28.

\bibitem[]{} 
Paczy\'nski, B., \& Wiita, P. 1980, A\&A, 88, 23.

\bibitem[]{} 
Phinney, E.~S., \& Kulkarni, S.~R. 1994, ARA\&A, 32, 591.

\bibitem[]{} 
Pringle, J.~E., \& Rees, M.~J. 1972, A\&A, 21, 1.

\bibitem[]{} 
Romani, R.~W. 1990, Nature, 347, 741.

\bibitem[]{} 
Shakura, N.~I., \& Sunyaev, R.~A. 1973, A\&A, 24, 337.

\bibitem[]{} 
Shu, F.~H., et al.~1994, ApJ, 429, 781.

\bibitem[]{} 
Spruit, H.~C., \& Taam, R.~E. 1990, A\&A, 229, 475.

\bibitem[]{} 
Stella, L., \& Vietri, M. 1997, submitted to ApJ (astro-ph/9709085).

\bibitem[]{} 
Stone, J.~M., \& Norman, M.~L. 1994, ApJ, 433, 746.

\bibitem[]{} 
Strohmayer, T., et al. 1996, ApJ, 469, L9.

\bibitem[]{} 
Sturrock, P.~A. 1991, ApJ, 380, 655.

\bibitem[]{} 
Urpin, V.~A., \& Geppert, U. 1995, MNRAS, 275, 1117.


\bibitem[]{} 
Van der Klis, M. 1997, in ``The Many Faces of Neutron Stars'' (Proc.
NATO ASI) (astro-ph/9710016).

\bibitem[]{} 
Van Kerkwijk, M.~H., van Paradijs, J., \& Zuiderwijk, E.~J. 1995, A\&A,
303, 497.

\bibitem[]{} 
Wang, J.~C.~L., Sulkanen, M.~E., \& Lovelace, R.~V.~E. 1990, ApJ, 355, 38.

\bibitem[]{} 
Wang, Y.-M. 1995, ApJ, 449, L153.

\bibitem[]{} 
White, N.~E., \& Zhang, W. 1997, ApJ, 490, L87.

\bibitem[]{} 
Wijnands, R.~A.~D., et al.~1997, ApJ, 479, L141.

\bibitem[]{} 
Yi, I. 1995, ApJ, 442, 768.

\bibitem[]{} 
Zhang, W., Strohmayer, T.~E., \& Swank, J.~H. 1997, ApJ, 482, L167
(astro-ph/9703151).

\end{thebibliography}
\end{document}